\documentclass[10pt,twocolumn,letterpaper]{article}
\usepackage[utf8]{inputenc}
\usepackage[english]{babel}
\usepackage[T1]{fontenc}
\usepackage{url}

\usepackage{textcomp} % for (TM) sign
\usepackage{graphicx}
\usepackage{abstract} % Allows abstract customization

\usepackage{algorithm}

\date{} % avoid date in \maketitle

\begin{document}

\title{The OCDF diagram. A metamodel for object-oriented systems visual design.}
\author{Alexander Reshytko \\
IBM Science and Technology Center, Moscow\\
{\tt\small areshytko@ru.ibm.com}}

\maketitle

\begin{abstract}

We present a metamodel for modeling control and data flows on subclass scales in object-oriented systems. UML Profiles were used as a representation mean and a complete metamodel definition was provided with an example of a diagram application.

\end{abstract}

\section{Introduction}

One can hardly deny the necessity of modeling in software engineering in the majority of product lifecycle phases along with related tasks that occur in this field - from designing a new system, to refactoring or even reverse engineering an existing one. The key concept of every model type is abstraction. We describe the system of interest in a very particular context from a very specific angle hiding all other aspects. Models thus make important implicit information evident, exlpicit. And more models we have for a system - more important characteristics of it are clear and visible to us. Another important thing to note here is that each and every model type is a language. And like every language it determines the way we think about the system. It determines the way we design them, the way we discuss them and the way we try to understand them.

This article deals with the modeling of object-oriented systems. In this field UML plays an important
role for over a decade now and it is highly promoted by OMG. It is an ISO standard and it has an abundant amount of literature about it as well as tools to support it. Probably the main reason for such monopoly is the fact that a wide variety of object-oriented systems aspects can be modeled by UML. We simply don't need anything else in most of the cases. However, there are limitations and certain aspects of object-oriented systems are only vaguely covered by traditional UML toolset. To address this, the new metamodel was developed - OCDF (object control-data-flow).

Let's begin by recapping some ideas behind object-oriented systems and their constituents.

\subsection{Object-oriented systems}

One of the common mathematical formalizations of an object-oriented methodology is via a notion of an Abstract Data Type (ADT)\cite{Meyer}. This concept is pretty self-explanatory. It is a data type hence it's incorporated in a programming language type system. And it is defined implicitly in a representation free form. It is achived via an ADT specification:

\begin{itemize}
	\item A type name (Types section);
	\item operations applicable to instances of the type given solely by their signature (Functions section);
	\item predicates which define function invariants (Axioms section);
	\item characteristic functions which define domains of partial functions (in case some of the functions listed in the section 2 are partial) (Preconditions section).
\end{itemize}

Now, a class is merely a tuple of three things: an ADT specification, a representation choice and a mapping from functions listed in the specification to the representation in the form of a set of mechanisms, or features, each implementing one of the functions in terms of the representation, so as to satisfy the axioms and preconditions. And an object-oriented software construction is the building of software systems as
structured collections of possibly partial abstract data type implementations.\cite{Meyer}

Class representation features can be devided into two separate kinds:

\begin{itemize}
	\item those, that are represented by space, that is to say by associating a certain piece of information with every instance of the class \cite{Meyer}. They are usually called attributes;
	\item those, that are represented by time, that is to say by defining a certain computation (an algorithm) applicable to all instances of the class \cite{Meyer}. They are usually called routines or methods.
\end{itemize}

This distinction is very important and concerns not only computation time / memory consumption dilemma. Attributes form the values space of class instances. They consitute the possibility for distinction between class instances. They are the reason why we talk about classes and their instances and not just collections of functions amended to a programming language's type system. Moreover, in the imperative paradigm attributes form a state of a class instance. This is how an instance can be changed during it's lifetime in most of the imperative object-oriented languages nowadays.

Now, if we think about kinds of relations that may occur between class representation feautes two most obvious candidates that come into mind are:

\begin{itemize}
	\item control-flow relaionship between routines - caller-callee relationship;
	\item data-flow relationship between routines or an attribute and a routine - suplier-provider relationship of a data in a form of function actual arguments, feature call result or a data value reading.
\end{itemize}

\subsection{Motivation}

We state that there are common use cases and situations when sub class structures formed by elements described above when modeled in a visual format can bring a lot of insights about the system being modeled. This is the idea behind OCDF models.

Ideally for the good granulated design with meaningfull naming conventions the static UML Class diagram is able to provide a good understanding of the system. Nevertheless, classes with a rather complex logic or numerous responsibilities (the God object antipattern is an extreme case) are a fairly common phenomenon nowadays. This is especially applicable to the long living legacy code. In other words, there is a practical task to support, refactor, fix, understand (and share this understanding) them. And metamodels that deal with inter-class relationship very often provide a very little help here.

One of the dangerous drawbacks of the locking-based threading model which is the main multithreading facility for many software engineering systems is a race condition possibility. And existing UML metamodels provide no facilities to cope with it.

Depicting loosly and strongly interconnected class substructes also helps in refactoring tasks and we found that there's often a functional reason behind such substructes which can be reflected in a more refined class system as a result of a refactoring process.

\section{OCDF diagram}

OCDF can help in such use cases and situations. In this metamodel a class is drawn as a directed graph of it's features (nodes) interconnected with control and data flows (arcs). It is a static diagram that is designed to accomplish legacy code understanding and support and refactoring tasks. It models the implementation level of a system.

What makes OCDF useful is it's ability to explicitly show tightly bound substructures of a class and boundaries between them. These are the boundaries that can be used to split the class into multiple ones during the refactoring process. These interconnected substructures fairly often embody different functional responsibilities and thus helps a reader (and an author) revealing such information.

Moreover we state that not only the model itself provides a good understandning of a class but the way the diagram is filled supports and structures it. It provides a gradual way of understanding a class inner logic producing artifacts incrementally as this understanding increases.

\section{The format of the metamodel}

We considered several meta-metamodeling frameworks as a tool to describe the OCDF metamodel: MOF, EMF, UML Profiles, and have chosen a latter one. This decision was determined by two factors: a simplicity of definition and an available tools support.

The formal specification of the OCDF metamodel defined as the UML Profile will now be given.

\section{The OCDF UML profile specification}

\subsection{Overview}

This profile describes the Object Control Data Flow (OCDF) metamodel for object-oriented systems. The profile is aimed to combine representational features of UML with Control Flow Diagram (CDF) and Data Flow Diagram (DFD).

A DFD is a graphical representation of the "flow" of data through an information system, modeling its process aspects. It can be viewed as a directed graph wherein the nodes are external entities, processes or data stores and the edges are data flows \cite{Bruza}.

Control flow differs from the Control Flow Graph and in a more high-level precision is just a caller-callee relationship between methods

The general idea behind OCDF models is to depict data and control flows in object-oriented systems on a sub-class scale. There are five basic metamodel elements to accomplish that:

\begin{itemize}
	\item Class Data member - represents the traditional OOP concept of a basic element of an object's state. It should have a one-to-one correspondence to modeled class's data members
	\item Class method - instances of this element in a model should have a one-to-one correspondence to modeled class's methods
	\item Data flow link - represents a directed data flow relationship between either a data member and a method or two methods.
	\item Control flow link - represents a directed control flow relationship between two methods.
	\item Class itself - basically a container for instances of all previously defined elements
\end{itemize}

Because of the nature of a UML profile these elements are defined as UML stereotypes which extend specific UML Superstructure metamodel elements. So they reuse and extend the standard UML semantics.

Although the most common use case is to describe a single class in the such type of diagrams there aren't any formal constraints and multiple classes can be modeled in a single diagram.

\subsection{Profile diagram}

\begin{figure*}
\centering
\includegraphics[width=1.8\columnwidth]{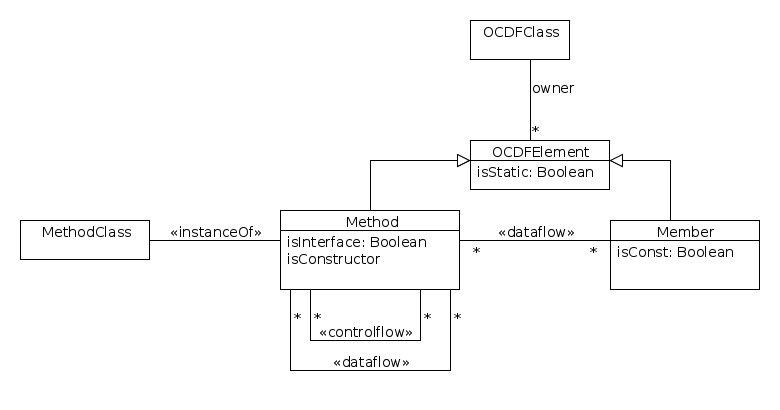}
\caption{
Informal structure
}
\label{fig:ocdf_logical}
\end{figure*}

\begin{figure*}
\centering
\includegraphics[width=2.2\columnwidth]{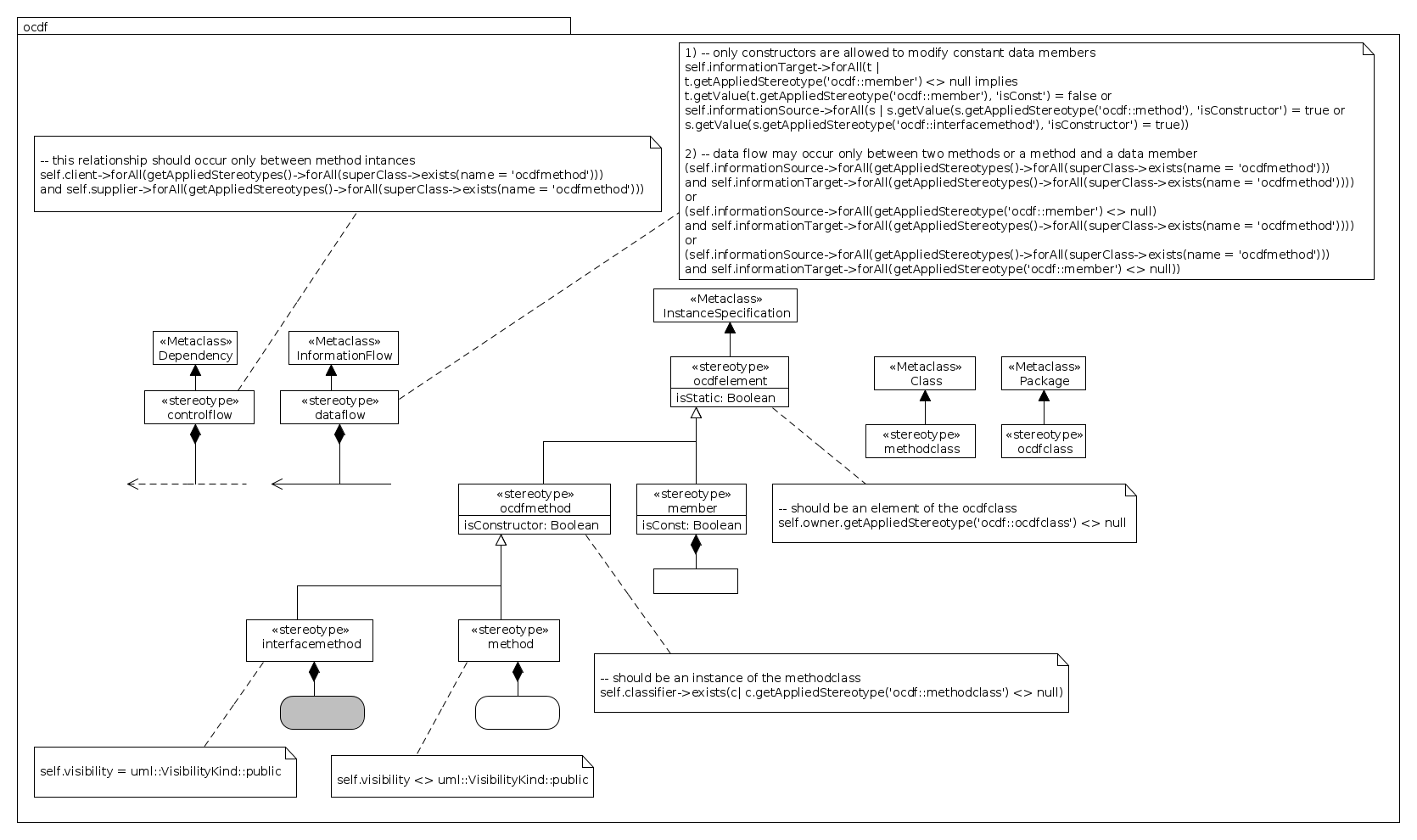}
\caption{
OCDF profile diagram
}
\label{fig:ocdf_profile}
\end{figure*}

Further are logical and formal diagrams for the metamodel described above. See Fig. \ref{fig:ocdf_logical} and Fig. \ref{fig:ocdf_profile}

\subsection{Stereotype Descriptions}

%===================================================================================================

\subsubsection{ControlFlow}

\paragraph{Generalizations:}

None.

\paragraph{Description:}

A ControlFlow is a special kind of a dependency relationship that may occur
between instances of a MethodClass and signifies a caller-callee relationship.

\paragraph{Attributes:}

None.

\paragraph{Extensions:}

Dependency (from Dependencies)

\paragraph{Constraints:}

[1] This relationship should occur only between method intances:
	\begin{flushleft}self.client->forAll(getAppliedStereotypes()->forAll(superClass->exists(name = 'ocdfmethod')))
	and self.supplier->forAll(getAppliedStereotypes()->forAll(superClass->exists(name = 'ocdfmethod')))\end{flushleft}

\paragraph{Semantics:}

The semantic of a control flow relationship is dependent on a programming language that implements a system being modeled and should mean a method call.

\paragraph{Notation:}

A control flow is shown as a dashed arrow between two method instances. The method instance at the tail of the arrow (caller) calls the method instance at the arrowhead (the callee). The arrow may be labeled with an optional name (as any dependency relation).

%===================================================================================================

\subsubsection{DataFlow}

\paragraph{Generalizations:}

None.

\paragraph{Description:}

A DataFlow is a kind of InformationFlow relationship that represents a directed data flow between either a data member and a method or two methods.

\paragraph{Attributes:}

None.

\paragraph{Extensions:}

InformationFlow (from InformationFlows)

\paragraph{Constraints:}

[1] Only constructors are allowed to modify constant data members
	\begin{flushleft}self.informationTarget->forAll(t | 
	t.getAppliedStereotype('ocdf::member') <> null implies
	t.getValue(t.getAppliedStereotype('ocdf::member'), 'isConst') = false or
	self.informationSource->forAll(s | s.getValue(s.getAppliedStereotype('ocdf::method'), 'isConstructor') = true or
	s.getValue(s.getAppliedStereotype('ocdf::interfacemethod'), 'isConstructor') = true))\end{flushleft}

[2] Data flow may occur only between two methods or a method and a data member
	\begin{flushleft}(self.informationSource->forAll(getAppliedStereotypes()->forAll(superClass->exists(name = 'ocdfmethod')))
	and self.informationTarget->forAll(getAppliedStereotypes()->forAll(superClass->exists(name = 'ocdfmethod'))))
	or
	(self.informationSource->forAll(getAppliedStereotype('ocdf::member') <> null)
	and self.informationTarget->forAll(getAppliedStereotypes()->forAll(superClass->exists(name = 'ocdfmethod'))))
	or
	(self.informationSource->forAll(getAppliedStereotypes()->forAll(superClass->exists(name = 'ocdfmethod')))
	and self.informationTarget->forAll(getAppliedStereotype('ocdf::member') <> null))\end{flushleft}

\paragraph{Semantics:}

The semantics is dependent on a programming language that implements
a system being modeled and should mean either of:
\begin{itemize}
	\item a data read of a data member by a method;
	\item a result of a method call;
	\item a data being passed as an argument to a method call.
\end{itemize}

\paragraph{Notation:}

A data flow is shown as an arrow between two OCDF elements. The element at the tail of the arrow signifies a
data provider and the element at the arrowhead signifies a data consumer.

%===================================================================================================

\subsubsection{InterfaceMethod}

\paragraph{Generalizations:}

OCDFMethod

\paragraph{Description:}

Represents an interface method of a modeled class.

\paragraph{Attributes:}

None.

\paragraph{Extensions:}

InstanceSpecification (from Kernel)

\paragraph{Constraints:}

[1] Should have a public visibility

	self.visibility = uml::VisibilityKind::public

\paragraph{Semantics:}

Should have a one-to-one correspondence to a modeled class's non-interface methods.

\paragraph{Notation:}

Is shown as a gray coloured rectangle with rounded corners, with the method signature shown inside the rectangle.

%===================================================================================================

\subsubsection{Member}

\paragraph{Generalizations:}

OCDFElement

\paragraph{Description:}

Represents a data feature (attribute) of a modeled class.

\paragraph{Attributes:}

isConst: Boolean

Constness of a data member means that it can't be changed once it's value is
defined in a constructor. But the semantic of this flag is generally dependent on a programming language that implements a modeled system

\paragraph{Extensions:}

InstanceSpecification (from Kernel)

\paragraph{Constraints:}

No additional consraints.

\paragraph{Semantics:}

Should have a one-to-one correspondence to modeled class's attributes.

\paragraph{Notation:}

Class data attributes are modeled as usual UML class instances - in a
rectangular box with a name, a visibility prefix and a type inside.

%===================================================================================================

\subsubsection{Method}

\paragraph{Generalizations:}

OCDFMethod

\paragraph{Description:}

Represents a non-interface method of a modeled class.

\paragraph{Attributes:}

None.

\paragraph{Extensions:}

InstanceSpecification (from Kernel)

\paragraph{Constraints:}

[1] Should have a non-public visibility

	self.visibility <> uml::VisibilityKind::public

\paragraph{Semantics:}

Should have a one-to-one correspondence to modeled class's non-interface
methods.

\paragraph{Notation:}

Is shown as a rectangle with rounded corners, with the method signature shown inside the rectangle.

%===================================================================================================

\subsubsection{MethodClass}

\paragraph{Generalizations:}

None.

\paragraph{Description:}

Represents a class of callable methods.

\paragraph{Attributes:}

None.

\paragraph{Extensions:}

Class (from Kernel)

\paragraph{Constraints:}

No additional constraints.

\paragraph{Semantics:}

Instances of this element in a model should have a one-to-one correspondence to a modeled class's methods.

%===================================================================================================

\subsubsection{OCDFClass}

\paragraph{Generalizations:}

None.

\paragraph{Description:}

Represents a class being modeled.

\paragraph{Attributes:}

None.

\paragraph{Extensions:}

Package (from Kernel)

\paragraph{Constraints:}

No additional constraints.

\paragraph{Semantics:}

Should have a one-to-one correspondence to a class in some object-oriented language.

%===================================================================================================

\subsubsection{OCDFElement}

\paragraph{Generalizations:}

None.

\paragraph{Description:}

Represents an element of the OCDFClass. Abstract.

\paragraph{Attributes:}

isStatic: Boolean

In some object-oriented languages there is a notion of a static class feature -
a feature of a class itself and not of it's instances. This boolean attribute
specifies this option.

\paragraph{Extensions:}

InstanceSpecification (from Kernel)

\paragraph{Constraints:}

[1] Should be an element of the ocdfclass.

	\begin{flushleft}self.owner.getAppliedStereotype('ocdf::ocdfclass') <> null\end{flushleft}

\paragraph{Semantics:}

This is an abstract metamodel element that represents a modeled class's features.

%===================================================================================================

\subsubsection{OCDFMethod}

\paragraph{Generalizations:}

OCDFElement

\paragraph{Description:}

Represents a callable feature (routine, method) of a modeled class. Abstract.

\paragraph{Attributes:}

isConstructor: Boolean

The notion of a constructor in many object-oriented languages means
a method used to create an instance of a class and usually has a separate call
syntax.

\paragraph{Extensions:}

InstanceSpecification (from Kernel)

\paragraph{Constraints:}

[1] Should be an instance of the MethodClass.

	\begin{flushleft}self.classifier->exists(c| c.getAppliedStereotype('ocdf::methodclass') <> null)\end{flushleft}

\subsection{Usage recommendations and suggestions}

As was already said the main purpose of this metamodel is to provide a help in understanding the legacy code, i.e. reverse-egnineer it. And so the diagram should accompany it's author in the process of investigating a class. Here a list of possible recommendations is provided:

\begin{itemize}
	\item The diagram has shown a good performance in combination with the UML Communication Diagram - for depicting both inter-class and in-class control flows.
	\item Modeling can be conveyed on different abstraction levels:
	\begin{itemize}
		\item just data flows between data members and methods
		\item data flow involving data members and control flow between methods
		\item the most detailized way involving all data flows between methods also together with with control flows (separate flow arrows for input parameters and output results) allowing for code generation and or obtained automatically from the source code (depicting all this information by hand might be tedious)
      \end{itemize}
	\item Inheritance requires a special treatment. We have found that a most convenient way is to depict parent features "lazily" by necessity - only if it's used in a current modeled child class explicitly. 
\end{itemize}

\section{An example of OCDF diagram}

\begin{figure*}
\centering
\includegraphics[width=2.2\columnwidth]{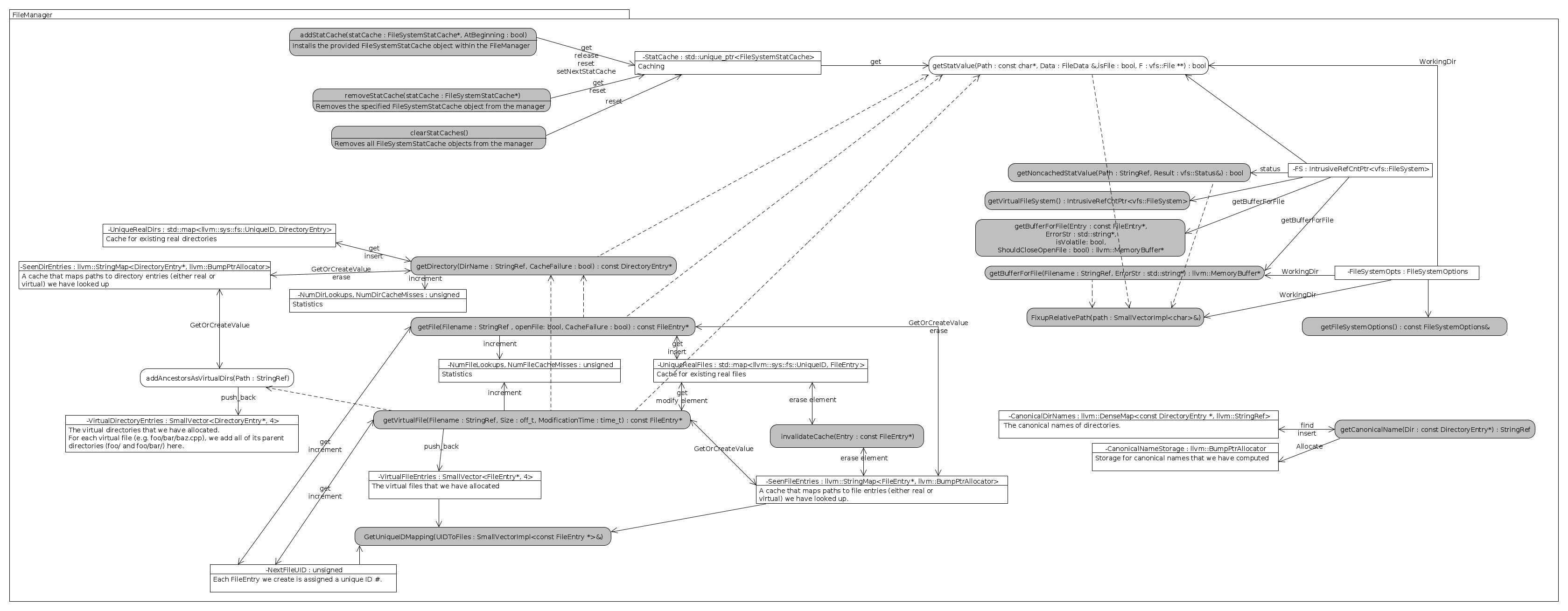}
\caption{
FileManager OCDF diagram
}
\label{fig:filemanager}
\end{figure*}

As an example of an OCDF diagram application we have chosen the FileManager class from the Clang project \cite{Clang}. The Subversion revision number is 212447. See Fig. \ref{fig:filemanager}

\section{Conclusions}

We introduced a new metamodel for modeling control and data flows on subclass scales in object-oriented systems. The UML Profiles were used to define the metamodel which means that OCDF diagram can be used by any tool that conforms UML 2 specifications.

\bibliographystyle{IEEEtran}

\end{document}